\documentclass[twocolumn,showpacs,preprintnumbers,amsmath,  amssymb,showkeys,floatfix,sort&compress,prl,longbibliography]{revtex4-1}
\usepackage{graphicx}
\usepackage{bmpsize}
\usepackage{bm}
\usepackage{cancel}
\usepackage{xcolor}
\usepackage{hyperref}
\hypersetup{
  colorlinks=true, 
  linkcolor=black,
  urlcolor=black,
  citecolor=black}
\usepackage{natbib}
\usepackage{booktabs}
\usepackage{xspace}

\newcommand{\mb}[1]{\mathbf{#1}}
\newcommand{\dt}{\partial_t}

\newcommand{\uv}{\mb{u}}
\newcommand{\zhat}{\mb{\hat{z}}}
\newcommand{\rhat}{\mb{\hat{r}}}
\newcommand{\lp}{\left(}
\newcommand{\rp}{\right)}

\begin{document}

\title{Subcritical thermal convection of liquid metals in a rapidly rotating sphere}
\author{E. J. Kaplan}
\affiliation{Univ. Grenoble Alpes, CNRS, ISTerre, F-38000 Grenoble}
\author{N. Schaeffer}
\affiliation{Univ. Grenoble Alpes, CNRS, ISTerre, F-38000 Grenoble}
\author{J. Vidal}
\affiliation{Univ. Grenoble Alpes, CNRS, ISTerre, F-38000 Grenoble}
\author{P. Cardin}
\affiliation{Univ. Grenoble Alpes, CNRS, ISTerre, F-38000 Grenoble}
\begin{abstract}
Planetary cores consist of liquid metals (low Prandtl number $Pr$)
that convect as the core cools. Here we study
nonlinear convection in a rotating (low Ekman number $Ek$) planetary
core using a fully 3D direct numerical simulation. Near the critical
thermal forcing (Rayleigh number $Ra$), convection onsets as thermal
Rossby waves, but as the $Ra$ increases, this state is superceded by
one dominated by advection. At moderate rotation, these states (here
called the weak branch and strong branch, respectively) are smoothly
connected. As the planetary core rotates faster, the smooth transition
is replaced by hysteresis cycles and subcriticality until the weak
branch disappears entirely and the strong branch onsets in a turbulent
state at $Ek < 10^{-6}$. Here the strong branch persists even as the
thermal forcing drops well below the linear onset of convection
($Ra=0.7Ra_{crit}$ in this study).
We highlight the importance of the Reynolds stress, which is required for convection to subsist below the linear onset.
In addition, the P\'eclet number is consistently above 10 in the strong branch.
We further note the presence of a strong zonal flow that is nonetheless unimportant to the convective state.
Our study suggests that, in the asymptotic regime of rapid rotation relevant for planetary interiors, thermal convection of liquid metals in a sphere onsets through a subcritical bifurcation.
\end{abstract}
\maketitle

Self-sustaining magnetic fields of terrestrial planets are 
generated in their liquid metal cores. Left on their own they would decay
from ohmic dissipation, but the slow cooling of the planets drives the
convective flows thought to maintain the fields \cite{Jones08}.
Numerical models of planetary dynamos \cite{Glatzmaier95,Miyagoshi10}
are widely used to study these strongly nonlinear systems. They solve
the Navier-Stokes equation coupled to a temperature equation and to
the induction equation that governs the magnetic field. For
simplicity, most direct numerical simulations (DNS) of the
dynamos have set the Prandtl number $Pr=1$. However, liquid metals
have $Pr \lesssim 0.1$ and the nature of their convection differs from
$Pr=1$ \cite{chandrasekhar1961,Guervilly.JFM.2016, calkins2016}.

The onset of convection in a full sphere is relevant to the early
history of the Earth's \cite{Olson.Science.2013} or Moon's
\cite{Laneuville2014} core, and has thus received a great deal of
attention ({\it e.g.}  \cite{Jones.JFM.2000}). A significant thermal
forcing is required to overcome the stabilizing rotational constraint
and drive convective instabilities\cite{Busse70,Jones.JFM.2000}. At
and near this threshold, convection onsets in the form of columnar
vortices aligned with the axis of rotation, drifting in the azimuthal
direction \cite{Busse70,Dormy04}. The nonlinear regime has been
extensively studied for $Pr=1$ \cite{gastine2016}.  The $Pr \ll 1$
regime is more difficult to tackle \cite{Zhang92}, but its nonlinear
state was recently described using a quasigeostrophic model, which
relies on a two dimensional description of the axial vorticity.  In
their simplified model, the authors of \cite{Guervilly.JFM.2016} found
first clues of subcritical convection---that is convection below the
linear onset of instability---anticipated by weakly nonlinear
theoretical predictions \cite{Soward77,Plaut08}.

%Because an imposed magnetic field can relax the strong Taylor constraint of rapid rotation \cite{chandrasekhar1961}, subcritical convection was expected in the case of strong self-sustained magnetic field (dynamo action). It was however never observed \cite{busse2011}.
%Without magnetic fields, it was only hypothesized that inertial (nonlinear) terms could lead to subcriticality \cite{Soward77,Plaut08}.

In the present work on the rotating convection problem, we use three-dimensional direct numerical simulations to describe the nature of the weak and strong convective branches, especially when the strong branch becomes subcritical.
We discuss the insights gained from the differences and similarities between the two branches.

\section{Formulation of the Model}

We study thermal Boussinesq convection driven by internal heating in a
sphere rotating at constant angular velocity \mbox{$\Omega \zhat$}.
The acceleration due to gravity is radial and increases linearly, as
expected for a constant density medium, \mbox{$\mb{g}=g_0 r
  \rhat$}. The radius of the sphere is $r_o$.  The fluid has kinematic
viscosity $\nu$, thermal diffusivity $\kappa$, density $\rho$, heat
capacity at constant pressure $C_p$, and thermal expansion coefficient
$\alpha$, all of which are constant.  We consider an homogeneous
internal volumetric heating S.%%   In the absence of convection, the
%% static temperature profile $T_s$ is calculated by solving the
%% diffusive heat equation and can be written as
%% \begin{equation}
%% 	T_s (r) = \frac{S}{6\kappa \rho C_p} (r_o^2-r^2) +T_o,\label{eq:Ts}
%% \end{equation}

The dynamic system is nondimensionalized by scaling lengths with
$r_o$, times with $r_o^2/\nu$, and temperature with \mbox{$\nu S
  r_o^2/(6\rho C_p\kappa^2)$}\cite{Guervilly.JFM.2016}. The system is governed by the
incompressible Navier-Stokes equation and an advection-diffusion
equation of the temperature perturbation,
\begin{align}
	\dt{\uv} + \left( \uv \cdot \nabla \right) \uv + \frac{2}{Ek}\zhat \times \uv 
	&= - \nabla p + \Delta \uv + Ra \Theta \mb{r} ,
	\label{eq:NS1}	\\
	\nabla \cdot \uv &= 0,	\\
	\dt{\Theta} + \uv \cdot \nabla \Theta -\frac{2}{Pr} r u_r &= \frac{1}{Pr} \Delta \Theta,
	\label{eq:T1}
\end{align}
\noindent where $\uv$ is the velocity field, $p$ is the modified
pressure, which includes the centrifugal potential, and $\Theta$ is
the temperature perturbation relative to $T_s$. The dimensionless
numbers are the Ekman number $Ek = \nu / r_o^2 \Omega$, the Prandtl number $Pr = \nu / \kappa$, and the Rayleigh
number $\lp Ra={\alpha g_0 S r_o^6} / {6 \rho C_p \nu \kappa^2}\rp$.
At $r=r_o$, the boundary condition for the velocity is no-slip and
impenetrable ($\uv = \mb{0}$) and the temperature is fixed ($\Theta = 0$).

This dynamic system is modeled in a 3 step process.  First, for each
set of $(Ek,Pr)$, the linear onset of convection $Ra_{crit}$ and the
associated eigenmode are computed precisely by using the
\texttt{singe} code \cite{Vidal.GJI.2015} to find the value of $Ra$
for which the least damped eigenmode of the linearized equations has
almost zero growth rate (see supplementary table \ref{tab:crit}). The
linear eigenmodes are then used as initial values of fully 3d
simulations, which are then run until they reach a statistical
equilibrium. Hysteresis cycles and subcriticality are explored by
changing the thermal forcing or rescaling the final flow state and
evolving the system to its new equilibrium.

The fully three-dimensional simulations are run with our spherical
code \texttt{xshells} \cite{schaeffer2013,marti2014}.
Most runs employ a form of hyper-viscosity
affecting only the 20\% highest spherical harmonics \cite{kaplan2016} to speed up computations, but we have checked that this does not
alter the solutions. 
Similarly, careful spatial and temporal convergence checks have been made, and we were surprised to find that the amplitude of the zonal winds (axisymmetric azimuthal flow) is very sensitive to the bulk radial grid spacing, although no sharp gradients were seen.
We suspect that the balance between the small viscous stress and the small Reynolds stress requires a high resolution to be computed accurately with second order finite differences.
As a result, simulations were run using up to 576 cores
for 1152 radial levels and spherical harmonics up to degree 199.

The simulations output several useful diagnostic values over the
course of the run. We present the velocity using the dimensionless
P\'eclet number ($Pe = U r_o/\kappa$) which is the ratio between the
rate of advection to the rate of diffusion of temperature. $U$ is
defined by the square root of the volume averaged kinetic energy; the
zonal P\'eclet number $\lp Pe_{zon}\rp$, uses only the azimuthal
component of the flow averaged in longitude ($m=0$); the convective
P\'eclet number $\lp Pe_{conv} \rp$ uses the $m\neq0$ velocity. We
also use the Nusselt number
$\left(Nu\equiv\frac{1}{Pr\Theta_c}\right)$ which measures the
convective heat transfer from the core to the surface. ($\Theta_c$ is
the temperature at the origin).

\section{Weak and Strong Branches}

Our studies seek to verify that the subcritical behavior found in the quasigeostrophic approximation \cite{Guervilly.JFM.2016} is also seen
in the fully three dimensional system, for $Ek \in \left[10^{-7}, 10^{-5}\right]$ and $Pr \in \left[ 0.003, 0.1\right]$.

\begin{figure}
  \includegraphics[width=\linewidth]{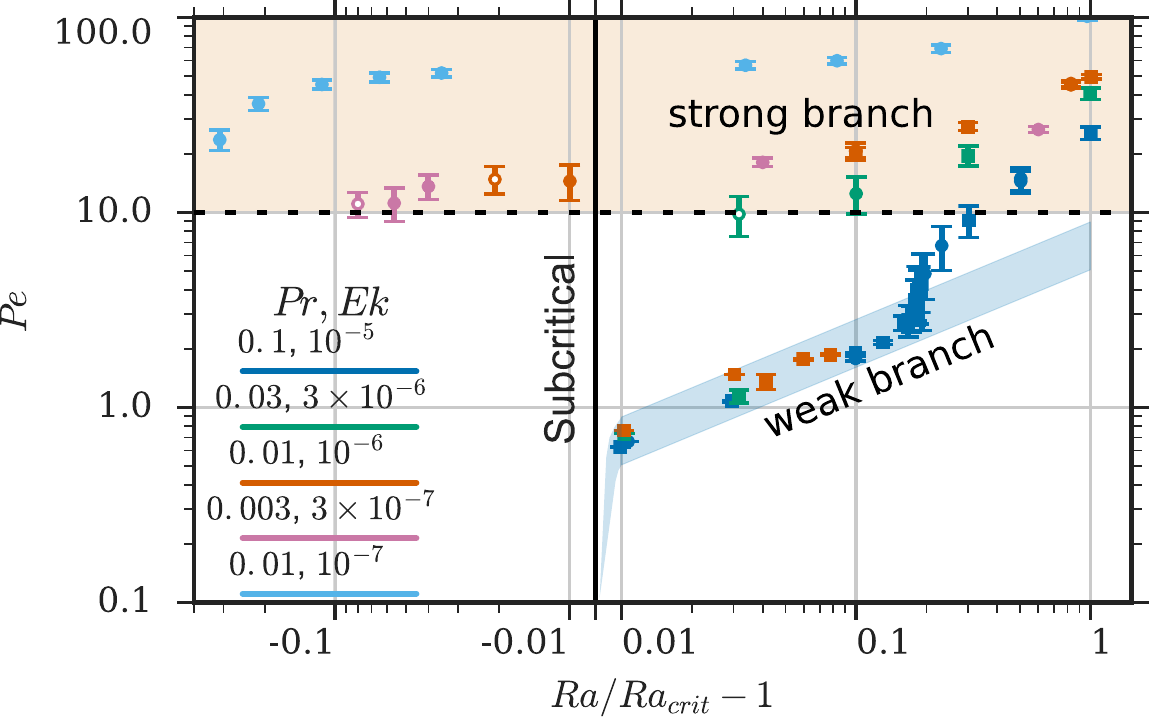}
  \caption{The mean velocity of the flows as a function of
    $Ra/Ra_{crit} - 1$. The $Ek$ and $Pr$ numbers are indicated by
    color. The error bars represent fluctuation levels. Open faced
    markers indicate simulations that were initialized from the strong
    branch that were observed to either transition to the weak branch,
    or decay to zero after a finite time greater than $\tau_\kappa$
    (the values and fluctuation levels are taken over the time before
    the transition or decay). The solid black line indicates the onset
    of the convective instability. The blue parallelogram, indicating
    the weak branch, scales as $Pe \propto \lp Ra-Ra_{crit}
    \rp^{1/2}$ \cite{Gillet06}.\label{fig:subcrit}}
\end{figure}

Our results are summarized in Fig.~\ref{fig:subcrit}, which shows the
P\'eclet number $\left(Pe\right)$ vs the Rayleigh number
$\left(Ra\right)$ for the simulations we've run.  Two branches are
clearly visible. a strong branch where the $Pe$ number is larger than
10 and a weak branch with lower $Pe$ numbers.
The advantage of representing the kinetic energies in terms of $Pe$ is that it
collapses all of the simulations onto a single scale with a clear divide between the strong and weak branches.
A more typical Reynolds number representation sees values of kinetic energy that are in either the weak branch or the strong branch depending on $Pr$; an example of this is in the supplementary figure \ref{fig:scalings}.
The $Pe$ is also important because one of the markers of the strong branch is a significant cooling of the sphere's core; $Pe > 10$ indicates that there is enough convective power to draw thermal energy away from the center.
At lower $Ek$, the strong branch persists below the linear critical
Rayleigh number.

\begin{figure}
  \includegraphics[width=\linewidth]{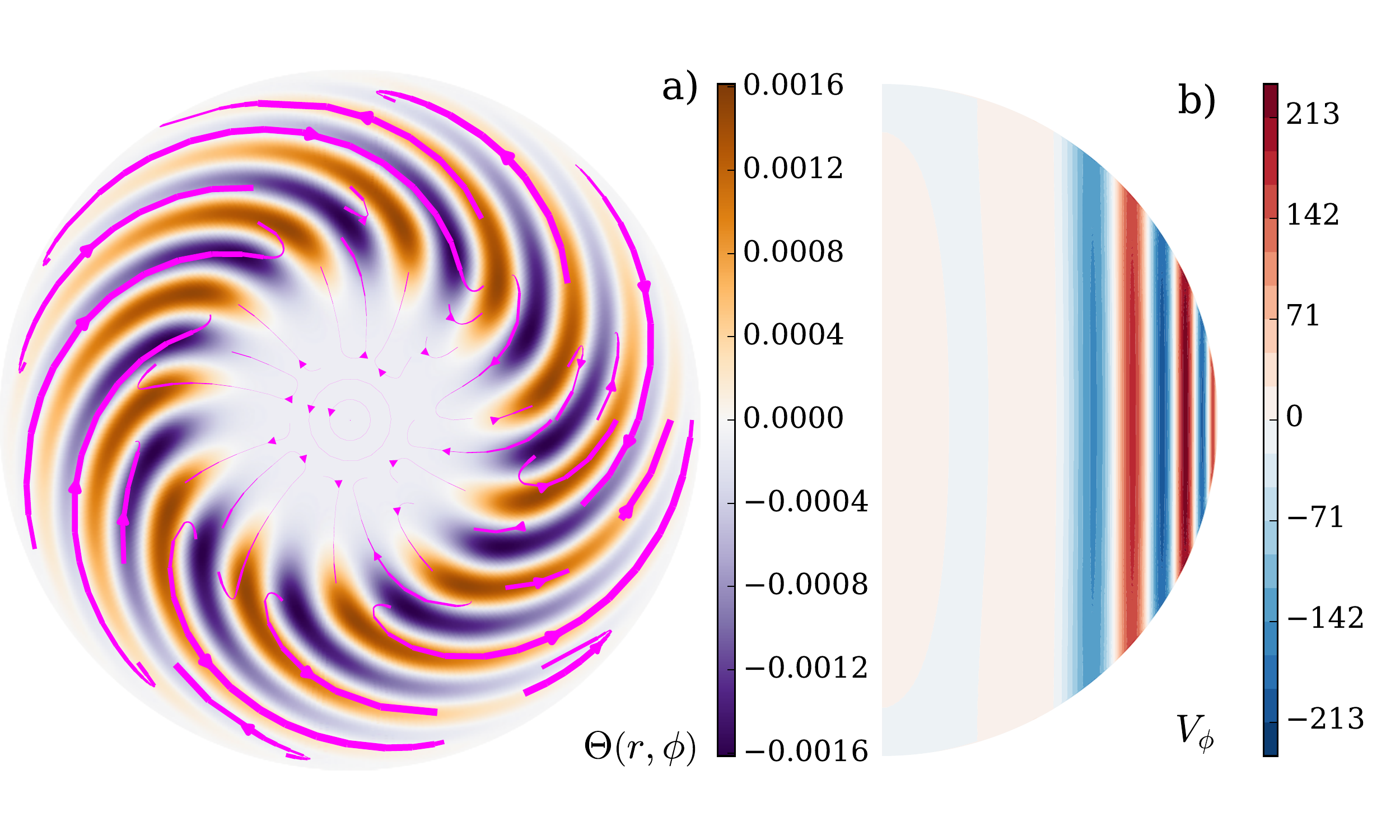}
  \caption{Cross sections of the weak branch system at $Ek=10^{-6}$,
    $Pr=0.01$, $Ra=5.53\times10^7=1.01Ra_{crit}$ showing (a) the temperature profile
    in the equatorial plane. Streamlines of the flow in the plane are
    plotted over the temperature profile in pink, and (b) Meridional
    slices of the azimuthal velocity. \label{fig:weakbranch_cuts}}
\end{figure}

The weak branch arises from a supercritical bifurcation at
$Ra_{crit}$. For Rayleigh numbers near this value, convection onsets as a thermal
Rossby wave \cite{Busse70}. The thermal anomaly $\Theta$ in the equatorial plane for
a typical case is visible in Fig.~\ref{fig:weakbranch_cuts}a \cite{Zhang92}. The
azimuthal velocity $v_\phi$ in a single meridional slice is shown in
Fig.~\ref{fig:weakbranch_cuts}b. The $z$ invariance implied by the
quasigeostrophic approximation is well displayed.
Near onset, the flow driven by the thermal gradient scales roughly with $\left(Ra - Ra_{crit}\right)^{1/2}$ \cite{Gillet06}, as shown in Fig.~\ref{fig:subcrit}.

\begin{figure}
  \includegraphics[width=\linewidth]{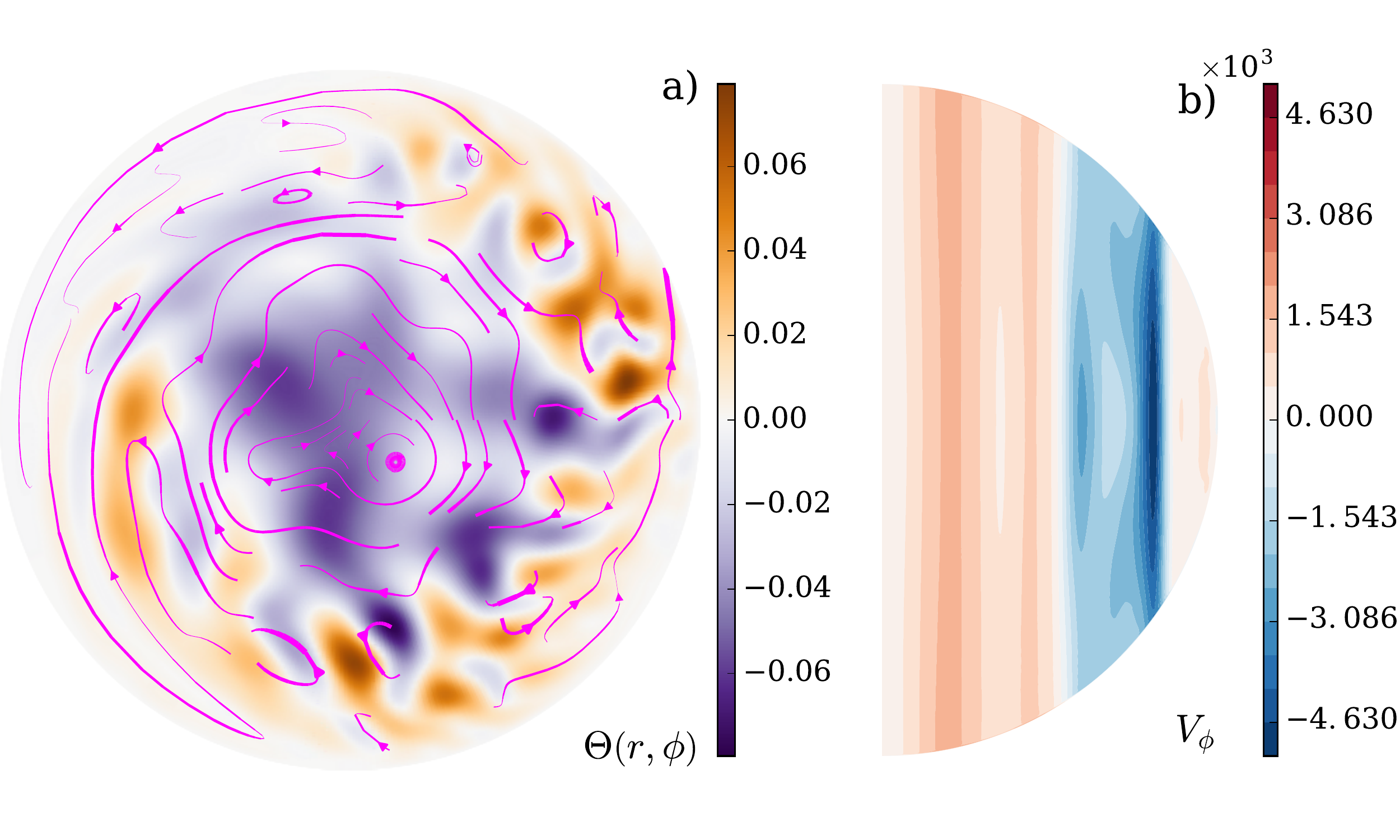} 
  \caption{Same as Fig.~\ref{fig:weakbranch_cuts}, but for the strong
    branch of the system at $Ek=10^{-6}, Pr=0.01,
    Ra=5.42\times10^7=0.99 Ra_{crit}$. \label{fig:strongbranch_cuts}}
\end{figure}

At large Rayleigh numbers, the weak branch becomes unstable
\cite{Guervilly.JFM.2016} and the flow evolves towards a state where
the velocity is an order of magnitude greater. A snapshot of a typical
strong branch flow is shown in Fig.~\ref{fig:strongbranch_cuts}. The
difference with Fig.~\ref{fig:weakbranch_cuts} is stark. There is
strong cooling near the origin of the sphere, and a noticeable
prograde zonal flow near the axis of rotation. There is some departure
from $z$ invariance, but the flow is still mostly columnar.

At $Ek=10^{-5}$ the strong branch is smoothly connected with the weak
branch. Below this $Ek$ the strong branch onsets with a discontinuity,
and is characterized by $Pe \gtrapprox 10$. This state onsets as a
subcritical bifurcation at $Ek \leq 3\times10^{-6}$, allowing a small hysteresis
cycle. The only green open-faced symbol in Fig.~\ref{fig:subcrit}
represents a simulation that persists in the strong branch for
$1.5\tau_\kappa$ ($\tau_\kappa=r_o^2/\kappa$ is the thermal diffusion time) before suddenly decaying to the weak branch. At $Ek=10^{-6}$ the
hysteresis persists below the onset of the weak branch instability.

At lower Ekman numbers ($Ek<10^{-6}$), the weak branch is absent and the thermal convection operates on the strong branch always at $Pe >10$.
Indeed, we have looked very close above the onset for $Ek=10^{-7}$ and $Pr=0.01$: even with an initial perturbation of very low amplitude ($Re \ll 1$) the kinetic energy quickly increased to reach the strong branch.
This demonstrates that the saturation mechanism leading to the weak branch at moderate Ekman numbers is lost at higher rotation rates.

\begin{figure}
  \includegraphics[width=\linewidth]{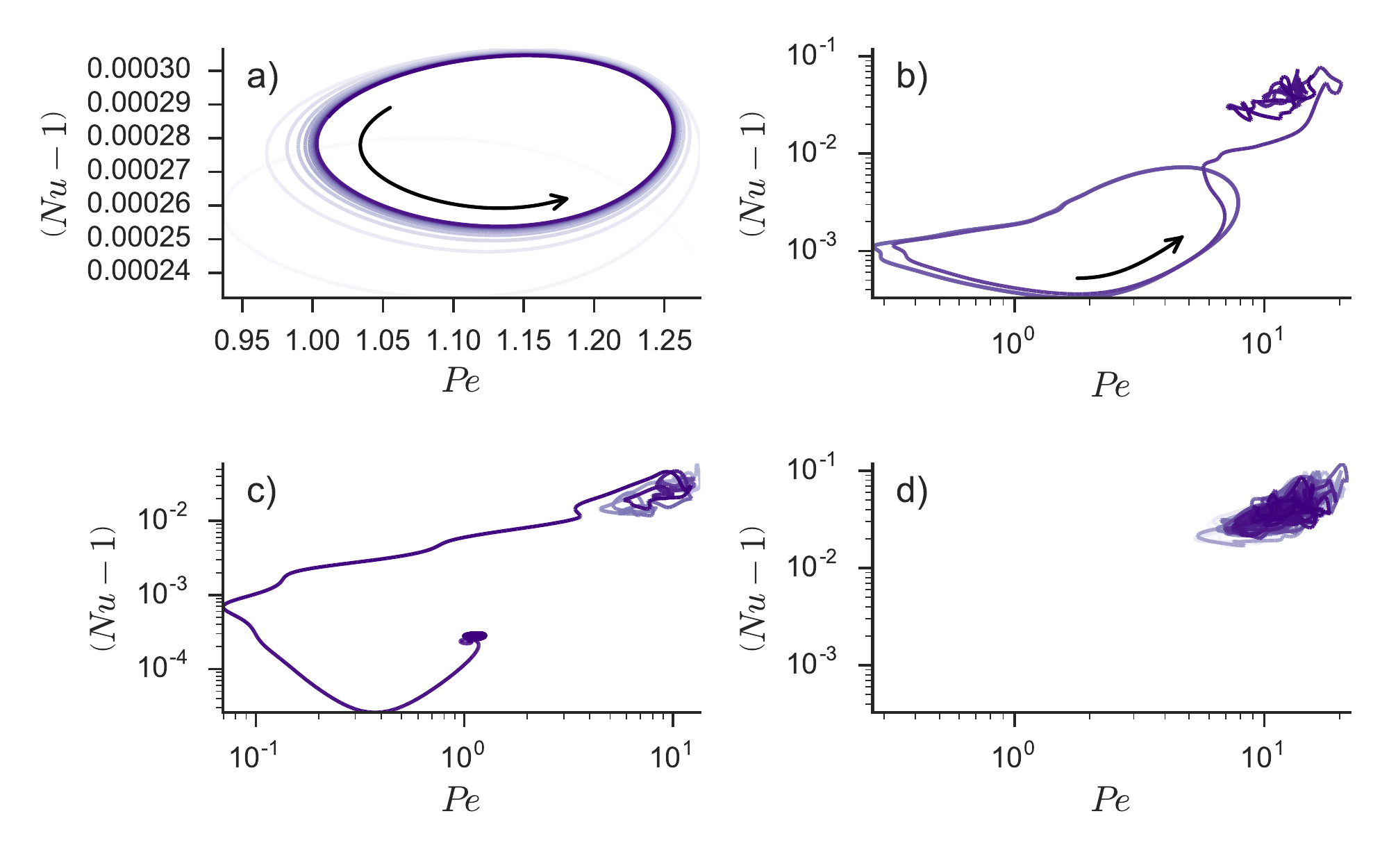}
  \caption{Phase trajectory of flow and core temperature of the flows
    for systems at $Pr=0.03, Ek=3\times10^{-6}$, and (a \& c) $Ra=1.03
    Ra_{crit}$ and (b \& d) $Ra=1.1 Ra_{crit}$. The top row column (a
    \& b) shows runs initialized from a weak branch state, the bottom
    row (c \& d) shows runs initialized from a strong branch
    state. The lines get darker as the system evolves. The arrows in
    (a \& b) indicate the direction of time. \label{fig:phasediag}}
\end{figure}

The phase trajectories of several simulations at $Pr=0.03,
Ek=3\times10^{-6}$ are shown in Fig.~\ref{fig:phasediag}. A simulation
at $Ra = 1.03Ra_{crit}$, initialized from a weak branch state is shown
in Fig.~\ref{fig:phasediag}a. The limit cycle is simple and stable; as
the convective power of the flow (indicated by $Pe$) increases, the
core cools (indicated by an increase in $Nu$), weakening the
convection, letting the core heat up again. Fig.~\ref{fig:phasediag}b
shows a simulation at $Ra = 1.1Ra_{crit}$, initialized from a weak
branch state. The limit cycle here is more complex.
%Rather than out of phase sinusoids, the convective cooling is strong enough to disrupt the convection.
The transition to the strong branch happens when other growing modes besides the critical one reach similar amplitude.
% though slowly enough that the weak branch persists over several $\tau_\kappa$.
Simulations at the same two Rayleigh numbers, but initialized in the
strong branch, are shown in Fig.~\ref{fig:phasediag}c~\&~d. Here we
see that the phase trajectories are stochastic rather than approaching
a limit cycle. At $Ra=1.03Ra_{crit}$, the system persists in the
strong branch for 1.5$\tau_\kappa$ before suddenly jumping to the weak
branch.
A second observation from Fig.~\ref{fig:subcrit} is that the
fluctuation levels in the strong branch actually decrease as the thermal forcing is increased.
These fluctuations are categorically different from the
fluctuations of the weak branch. Here, the fluctuations are stochastic
in the phase spaces of the various diagnostic parameters. 

\section{Subcriticality}
As Fig.~\ref{fig:subcrit} shows, subcritical convection exist in our
system, as convection on the strong branch can occur below the
linear onset of convection, at $Ra < Ra_{crit}$.  Interestingly at
$E/Pr = 10^{-4}$ the subcriticality is small and fragile: a small
perturbation can kick the system back to rest. This is indicated by
the open symbols below $Ra < Ra_{crit}$ (Fig.~\ref{fig:subcrit}),
where the flow transitions to rest because of its intrinsic
fluctuations, sometimes after several thermal diffusion times.

A much more robust subcriticality is observed at $E/Pr = 10^{-5}$ and
$Pr=0.01$.
%Note that these simulations are demanding: 1152 radial grid points
%and 199 spherical harmonic degree were needed for a good convergence
%of the zonal flow strength.
There, the strong branch can be violently perturbed without loosing
the convection.  As an example, we can divide flow and temperature
anomalies by 10 at $Ra = 0.94 Ra_{crit}$ and the convection will quickly
recover, while dividing instead by 100 sends the system back to rest.
By gradually lowering $Ra$, we have observed robust convection at
decreasing $Ra$ down to $Ra = 0.69 Ra_{crit}$ and lost the convection at
$Ra = 0.64 Ra_{crit}$, when the P\'eclet number dropped below $Pe \simeq
10$.  Further stress tests were applied at $Ra = 0.94 Ra_{crit}$.  We
artificially removed all $m=0$ (axisymmetric) components, including
the zonal flow and zonal temperature anomaly: the convection stayed
firm.  We also kept only the most unstable mode at onset and its
harmonics: the convection endured.
When removing the $(\uv.\nabla)\Theta$ term, the kinetic energy increases significantly.
Instead, when removing the $(\uv.\nabla)\uv$ term, the subcritical convection dies out.

In the 2D planar geometry studied by Chandrasekhar \cite{chandrasekhar1961}, Veronis predicted a possible subcriticality in a window of low rotation rates \cite{veronis1959}, recently confirmed numerically \cite{beaume2013}.
The subcritical behavior is associated with the presence of large mean flows which reduce locally the effective rate of rotation and consequently, the rotational constraint on the flow and the critical Rayleigh number toward its non rotating value.
This mechanism implies a Rossby number $Ro$ close to unity.
In contrast, it is very small here ($Ro = Re.Ek <10^{-3}$) because of the high rotation rate ($Ek=10^{-7}$) and despite the large Reynolds numbers ($Re \sim 10^4$) observed near the convection onset.
It is thus not surprising that the zonal flow is not important for subcritical convection.
Furthermore, outside the boundary layers and in the subcritical regime, the root-mean-square local vorticity fluctuations never exceed $5\%$ of the background vorticity $2\Omega$.
% moreover, the local voriticity seems to decrease as the Ekman number is decreased ? yes or no ?
Not only did Veronis find no subcritical motion in this rapidly rotating regime \cite{veronis1968}, but the mechanism of lowering the effective (local) rotation rate cannot explain the large amount of subcriticality we found (convection down to $Ra = 0.69 Ra_{crit}$).

Our numerical experiments highlight the key role of the Reynolds stress to sustain convection below the linear onset.
However, it is the P\'eclet number that is more conveniently used to characterize the strong branch and subcriticality.
If $Pe$ is much larger than $10$ at the linear onset
($Pe(Ra_{crit}) \simeq 50$), it is possible to have convection at $Ra$ well below $Ra_{crit}$, as long as $Pe \gtrsim 10$.
Below that threshold, the strong branch of convection cannot survive.

Our study also shows the trend of lower and lower subcritical $Ra/Ra_{crit}$ as $Ek/Pr$ is reduced while keeping $Pr \sim 0.01$.
This suggests an even larger effect at planetary core conditions ($Ek/Pr < 10^{-10}$), currently out of reach for numerical models.

\section{Conclusions}
\begin{figure}
  \includegraphics[width=\linewidth]{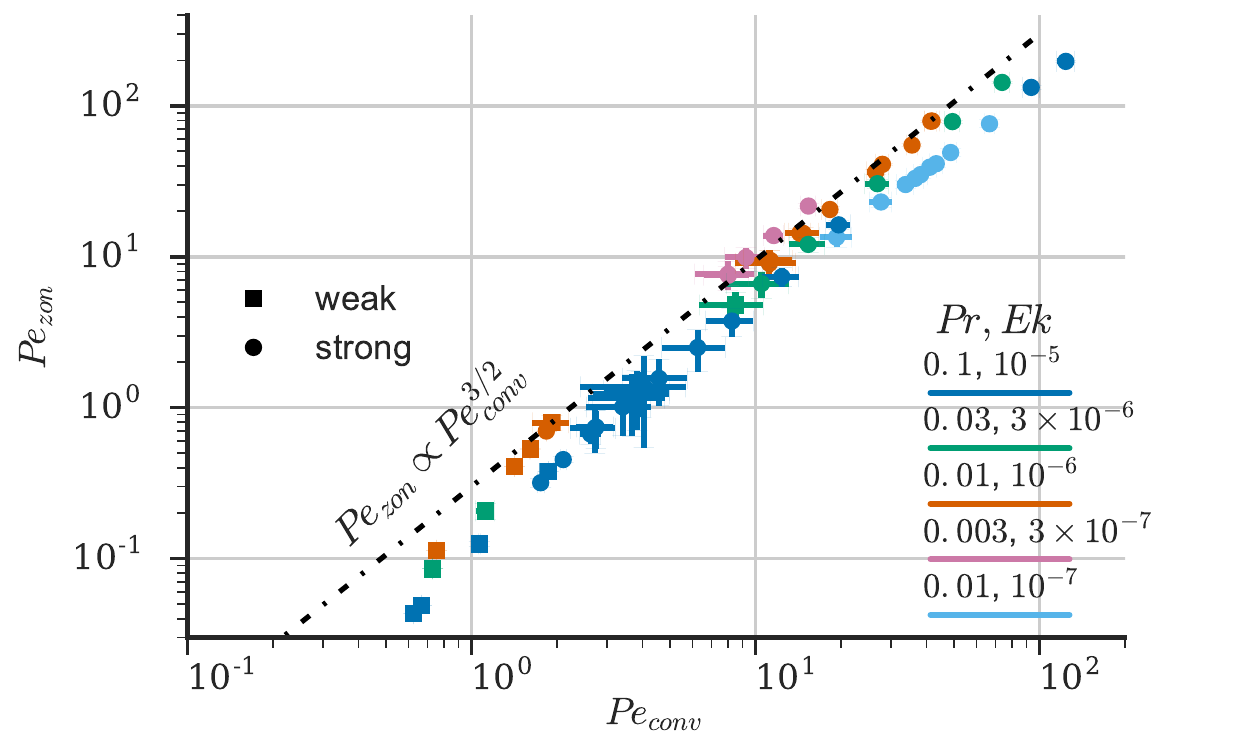}
  \caption{The averaged zonal velocity $Pe_{zon}$ of the flows as a function of the averaged convective velocity $Pe_{conv}$. The $Ek$ and $Pr$ numbers are indicated by color. The shape
    of the markers indicate the branch the averages were taken
    over. The error bars represent fluctuation levels. 
    \label{fig:strongscaling}}
\end{figure}

Subcriticality, like hysteresis, implies the presence of active
nonlinearities. The zonal flow in this system is produced by nonlinear
interaction of the convective velocity (Reynolds stress). In
Fig.~\ref{fig:strongscaling} we plot the $Pe_{zon}$ as a function of
$Pe_{conv}$ over the full set of simulations carried out. The data
points seem to align to an inertial scaling law with a power of 3/2
\cite{Gillet06}. This scaling law persists not only between the
different $Ek$ and $Pr$ numbers, but between the strong and weak
branches as well. A set of subcritical runs were rerun with their
zonal components artificially canceled (set to zero at every time
step) with no significant change in the convective flow or core
temperatures.  These two factors combined show that the zonal flow is
only ever a byproduct of the convective flows and not a driver of any
of the dynamics.
This is fundamentally different from the subcriticality predicted for moderate Ekman numbers ($Ek \sim 1$), for which the mean flow weakens the stabilizing effect of global rotation \cite{veronis1959,beaume2013}.

\section{Acknowledgements}
We thank three anonymous reviewers for their constructive comments.
The SHTns, SINGE and XSHELLS codes are freely available at \url{https://bitbucket.org/nschaeff/}.
We acknowledge GENCI for awarding us access to resource Occigen (CINES) and Turing (IDRIS) under grant x2015047382 and x2016047382.
Part of the computations were also performed on the Froggy platform of CIMENT (\texttt{https://ciment.ujf-grenoble.fr}), supported by the Rh\^ one-Alpes region (CPER07\_13 CIRA), OSUG@2020 LabEx (ANR10 LABX56) and Equip@Meso (ANR10 EQPX-29-01).
This work was partially supported by the French {\it Agence Nationale de la Recherche} under grants ANR-13-BS06-0010 (TuDy) and ANR-14-CE33-0012 (MagLune).

\clearpage

\appendix
\section{Supplementary Materials}

% reset figure counter and number with "S"
\setcounter{figure}{0}
\makeatletter 
\renewcommand{\thefigure}{S\@arabic\c@figure}
\makeatother

\subsection{Linear results}
The critical parameters of the linear onset of convection shown in table \ref{tab:crit} have been determined using the freely available \texttt{singe} code \cite{Vidal.GJI.2015}, which has been benchmarked with the results of Jones et al \cite{Jones.JFM.2000}.

\begin{table}
\caption{\label{tab:crit} Critical parameters at the onset of convection}
\begin{ruledtabular}
\begin{tabular}{llccc}
$Ek$ & $Pr$ & $Ra_{crit}$ & $m_c$ & $\omega_c/\Omega$\\
\hline
$10^{-5}$ & $0.1$ & $8.440 \times 10^6$ & 11 & $-0.04024$  \\ %by JV
$3 \times 10^{-6}$ & $0.03$ & $2.336 \times 10^7$ & 12 & $-0.04275$  \\  %by NS
$10^{-6}$ & $0.01$ & $5.475 \times 10^7$ & 11 & $-0.03895$  \\  %by NS
$3 \times 10^{-7}$ & $0.003$ & $1.255 \times 10^8$ & 12 & $-0.04287$ \\ %by NS
\hline
$10^{-7}$ & $0.01$ & $1.01574 \times 10^9$ & 27 & $-0.01930$  \\ %by NS
$10^{-7}$ & $0.01$ & $1.01567 \times 10^9$ & 28 & $-0.01941$  \\ %by NS
\end{tabular}
\end{ruledtabular}
\end{table}

For $Ek=10^{-7}$, $Pr=0.01$, the onset of modes $m=27$ and $m=28$ are very close to each-other, and we list both for completeness.

\subsection{Complementary nonlinear results}

\begin{figure}
  \includegraphics[width=\linewidth]{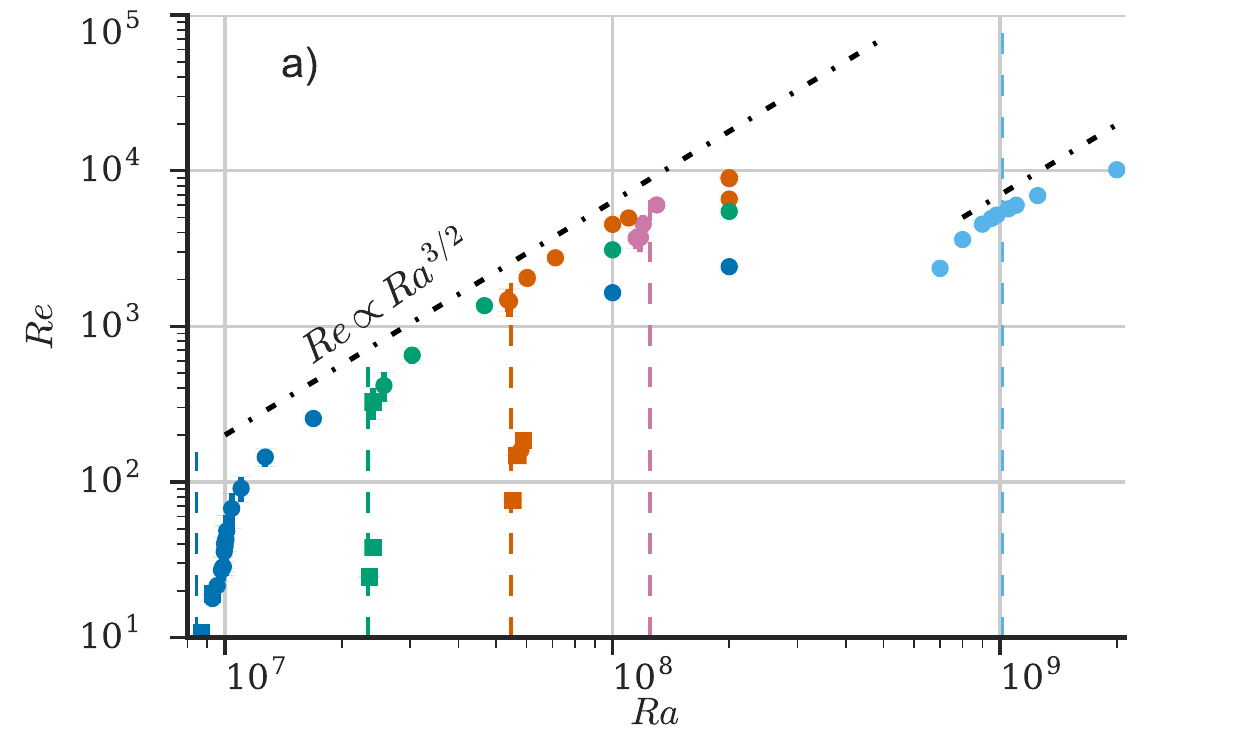} \\
  \includegraphics[width=\linewidth]{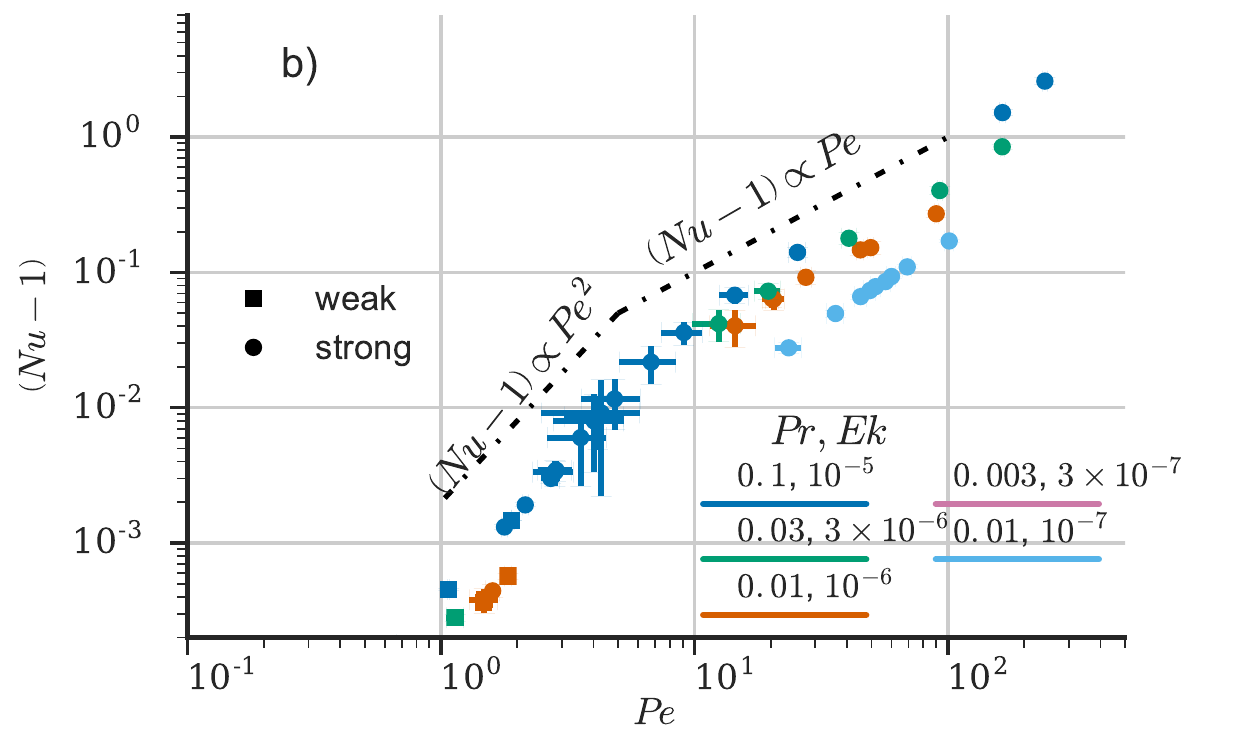}
  \caption{a) The mean velocity of the flows as a function of
    $Ra$. The $Ek$ and $Pr$ numbers are indicated by color. The shape
    of the markers indicate the branch the averages were taken
    over. The error bars represent fluctuation levels. Vertical lines
    indicate the critical Rayleigh number $Ra_{crit}$ for each $\lp
    Ek, Pr \rp$. The error bars represent fluctuation levels, in most
    cases these are more than an order of magnitude smaller than the
    mean level, and thus not visible in the figure.  b) The $\lp
    Nu-1\rp$ plotted against the mean $Pe$ of the system.  }
    \label{fig:scalings}
\end{figure}

A more typical measure of the convective velocity is the Reynolds
number ($Re = U r_o/\nu$) which is in our case related to the P\'eclet
number by $Re = Pe/Pr$. Fig.~\ref{fig:scalings}a shows the convective
velocity versus thermal forcings $Ra$. Near the linear onset of
convection, the strong branch flows seem to align with a single
trendline for each ratio $Ek/Pr$ that seems to scale with $Ra^{3/2}$. As the
$Ra$ increases, the flows depart from this 'scaling law', most likely
as a result of Ekman pumping.  Fig.~\ref{fig:scalings}b shows the
convective heat transfer as a function of convective velocity.  Large
forcings (strong branch) seems to get a lower exponent than the 2
exponent expected by a weakly non linear analysis at the onset of
convection \cite{Gillet06}. These trendlines should be seen as
suggestions to guide the eyes, rather than hard and fast scaling laws
that the data must conform to.

All the data behind Figs.~\ref{fig:strongscaling}~\&~\ref{fig:scalings}
and a jupyter notebook reproducing the figures is available at
\url{https://dx.doi.org/10.6084/m9.figshare.4540846} for anybody to
plot data as needed.

\begin{figure}
  \includegraphics[width=\linewidth]{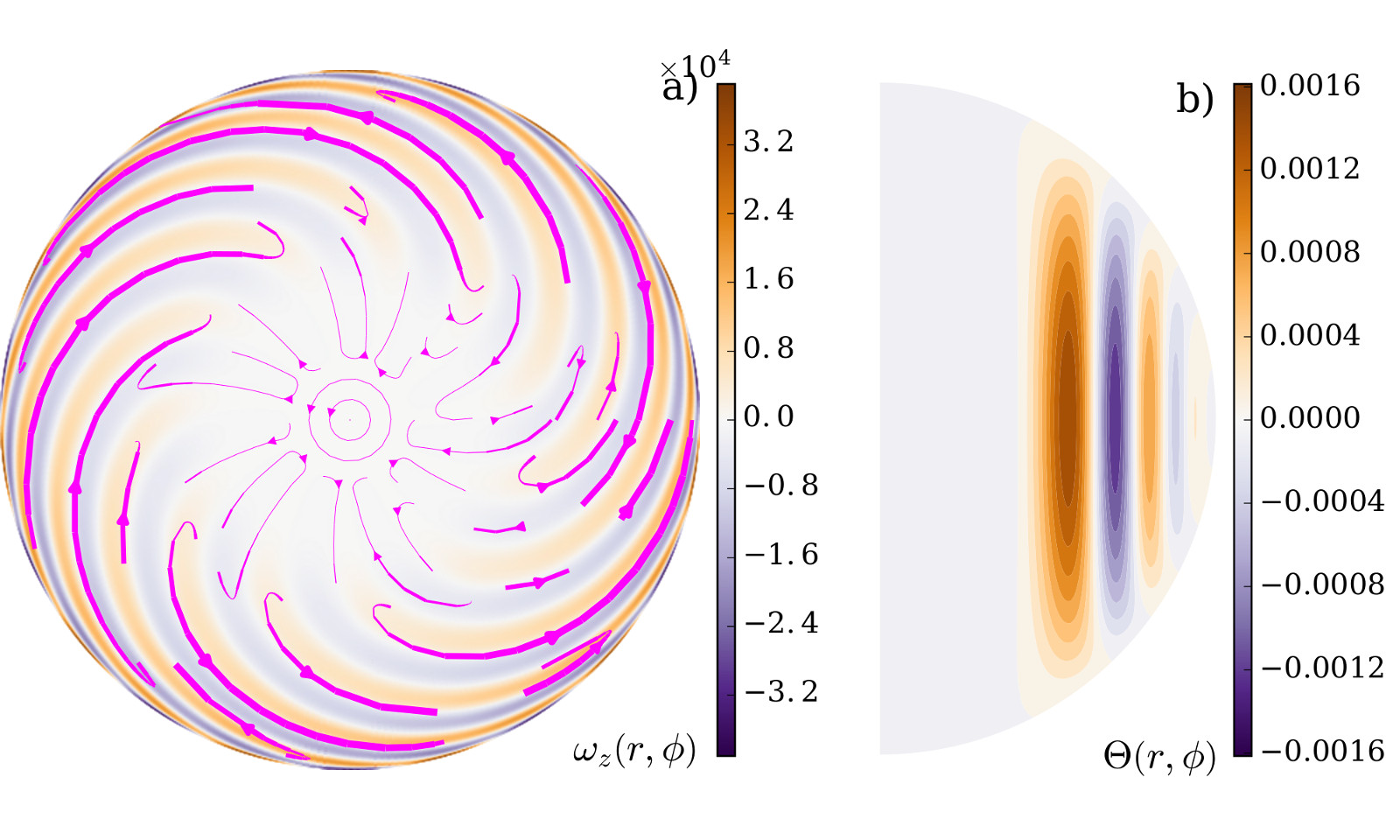}
  \caption{Cross sections of the weak branch system at $Ek=10^{-6}$,
    $Pr=0.01$, $Ra=5.53\times10^7=1.01Ra_{crit}$ showing (a) the vertical vorticity profile  in the equatorial plane. Streamlines of the flow in the plane are
    plotted over the vorticity profile in pink, and (b) Meridional slices of the thermal anomaly. These complement figure \ref{fig:weakbranch_cuts}.
    \label{fig:weakbranch_cuts_alt}}
\end{figure}

\begin{figure}
  \includegraphics[width=\linewidth]{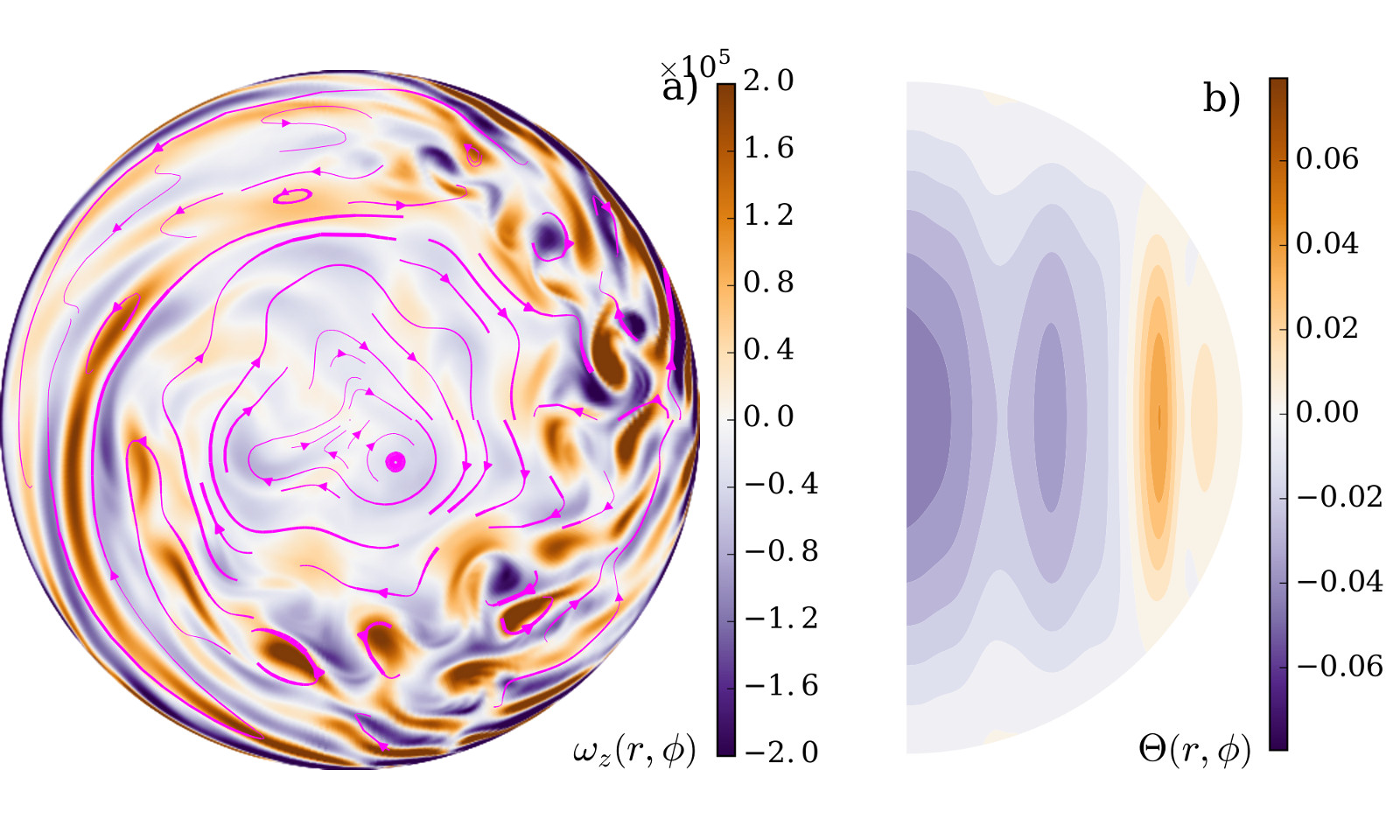} 
  \caption{Same as Fig.~\ref{fig:weakbranch_cuts_alt}, but for the strong
    branch of the system at $Ek=10^{-6}, Pr=0.01,
    Ra=5.42\times10^7=0.99 Ra_{crit}$. These complement figure \ref{fig:strongbranch_cuts}.
    \label{fig:strongbranch_cuts_alt}}
\end{figure}

Finally, to complement the fields shown in figure \ref{fig:weakbranch_cuts} and \ref{fig:strongbranch_cuts}, we represent thermal anomaly in the meridional plane and vertical vorticity in the equatorial plane in figure \ref{fig:weakbranch_cuts_alt} and \ref{fig:strongbranch_cuts_alt}.

\clearpage
\end{document}